\documentstyle[12pt]{article}

\newcommand{\A}{{\bar A}}

\def\bfcalD{\mbox{\boldmath${\cal D}$}}
\newcommand \beq{\begin{eqnarray}}
\newcommand \eeq{\end{eqnarray}}
\def\bfgrad{\mbox{\boldmath$\grad$}}

\def\del{\partial}    
\def\grad{\nabla}                               
\begin{document} 

\title{ 
Variational calculations in gauge theories with approximate projection on
gauge invariant states \\
}

\author{C. Heinemann, C. Martin, D. Vautherin\\
Division de Physique Th\'eorique \footnote{Unit\'e de Recherche 
des Universit\'es de Paris XI et Paris VI associ\'ee au C.N.R.S.}\\
Institut de Physique Nucl\'eaire\\
F--91406 Orsay Cedex, France\\
and\\
\\
E. Iancu\\
SPhT, CEA Saclay, F91191- Gif sur Yvette Cedex, France}

\maketitle

\begin{abstract}
Variational calculations using Gaussian wave functionals combined with 
an approximate projection on gauge invariant states are presented. 
We find that the energy exhibits a minimum for a wave functional centered
around a non vanishing background magnetic field. We show that divergences
can be removed by a renormalization of the coupling constant. The resulting 
expectation value of the gluon condensate is found to be in qualitative
agreement with phenomenological estimates.

\end{abstract}

\date{Preprint IPNO/TH 97-10, SPhT 97/00}

\newpage
The application of the variational method to gauge theories is generally
plagued by the difficulty to implement in a calculable way the requirement
of gauge invariance of physical states \cite{GOGNY,KERMAN,KOVNER}. In the
present letter we propose to use an approximate projection method developped
in 1962 by Thouless and Valatin \cite{THOULESS}. This method deals with the
restoration of rotational invariance when deformed solutions are obtained in
nuclear Hartree-Fock calculations.

In the case of gauge theories this formalism allows one to explore a class
of invariant trial vacuum functionals for which closed expressions can be
obtained.This class is generated by projecting Gaussian wave functionals
onto the subspace of gauge invariant states. This class of trial
functionals has already been explored by Kogan and Kovner \cite{KOVNER}, and
Brown and Kogan \cite{BROWN}.
Since these authors use different approximation schemes their approach and
ours appear complementary.

We consider the functional Schr\"odinger description of
the SU($N$) Yang-Mills theory. In the temporal gauge $A^0_a=0$,
the canonical coordinates are the vector 
potentials $A^i_a({\bf x})$ and the electric fields $E^i_a({\bf x})$,
which we shall often write as color matrices in the adjoint
representation: e.g., $A^i\equiv A^i_b T^b$
(the color indices $a, \,b,\dots$ run from 1 to $M\equiv N^2-1$).
The generators $T^a$ of the color group are taken to be
Hermitian and traceless; they satisfy \beq\label{lie}
[T^a,T^b]=if^{abc}T^c,\qquad\,\,\,\,
{\rm Tr}\,(T^a T^b)=N\delta^{ab},\qquad\,\,\,\,
 (T^a)_{bc}=-if^{abc}.\eeq
The  Hamiltonian density reads
\beq\label{HTG}
{\cal H}({\bf x})\,=\,\frac{1}{2}\left\{E^i_a E^i_a({\bf x})
\,+\,
B^i_a B^i_{a}({\bf x})\right\},\eeq
with the color magnetic fields $B^{i}_a\equiv -\epsilon^{ijk} F^{jk}_a/2$,
and $F^{ij}_a\equiv \del^i A^j_a - \del^j A^i_a + g f_{abc} A^i_b A^j_c$
($g$ denotes the coupling constant).
By also introducing the covariant derivative $D^i\equiv\del^i - i g A^i$,
we can rewrite the previous formulae in matrix form: e.g.,
$ F^{ij} \equiv F^{ij}_a T^a = [D^i, D^j]/(ig)$.

In the Schr\"odinger representation, the states
are represented by functionals of $A^i_a({\bf x})$,
$\Psi[{\bf A}]$, and the electric field is acting
on such states by functional differentiation:
\beq
 E^i_a({\bf x}) \Psi[{\bf A}]\,=\, i\,\frac{\delta}{\delta A^i_a({\bf x})}
\,\Psi[{\bf A}]\,.\eeq
The hamiltonian $H$ commutes
with the generator ${\cal G}$ of time-independent gauge transformations,
\beq\label{GL}
{\cal G}({\bf x})\,\equiv\,\bfgrad \cdot {\bf E}({\bf x})
\,+\,ig[A^i,E^i],\eeq
so it is possible to diagonalize $H$ and ${\cal G}$ simultaneously.
We are here interested only in those eigenstates $\Psi$ of $H$ which are gauge
invariant, i.e.,
\begin{equation} \label{GAUSS}
{\cal G}({\bf x}) \Psi[{\bf A}]\,=\,0,
\end{equation}
or in superpositions of them, to be referred to as {\it physical states}.

The variational principle consists in the inequality
\begin{equation} \label{Ritz}
\langle H \rangle \equiv 
\frac{< \Psi | H | \Psi >}{< \Psi | \Psi >}
\ge E_{vac},
\end{equation}
which holds for any functional $\Psi[{\bf A}]$ from the
physical Hilbert space. In this equation, $E_{vac}$ is 
the ground state energy, assumed to be non-degenerate.
In practice, however, one has to restrict oneself
to {\it Gaussian} wave functionals, the only ones 
which allow for the analytical computation of $\langle H \rangle$.
These have the form
\beq\label{GTWF}
\Psi[{\bf A}]\,=\,{\rm exp}\left\{-\,\frac{1}{4}
\int {\rm d}^3 x {\rm d}^3 y\left[A^i_a({\bf x})
- \bar A^i_a({\bf x})\right]\left(G^{-1}\right)^{ab}_{ij}
({\bf x, y})\left[A^j_b({\bf y})
- \bar A^j_b({\bf y})\right]\right\},\eeq
where the background field ${\bar A}^i_a({\bf x})$ and the kernel
$G^{-1}$ (with matrix elements $\left(G^{-1}\right)^{ab}_{ij}
({\bf x, y})$) are the variational parameters.

The expectation value of the Hamiltonian density in the Gaussian state
$\Psi$ is \cite{KERMAN}
\begin{equation} \label{ENERGY}
\begin{array}{lll}
\langle \Psi|{\cal H}({\bf x})| \Psi \rangle&=&
\frac{1}{2} {\bar {\bf B}}\cdot {\bar {\bf B}} ({\bf x})
\,+\,\frac{1}{8}{\rm Tr}\,\langle {\bf x}|G^{-1}|{\bf x}\rangle 
\,+\,\frac{1}{2} {\rm Tr}\, \langle {\bf x}|K G|{\bf x}\rangle\\
&&
\\
&&
+\,\frac{g^2}{8}\,
\left( {\rm Tr}\, \{ S_i T^a \langle {\bf x}|G|{\bf x}\rangle
\} \right)^2 \\
&&
\\
&&
+\,\frac{g^2}{4}\,{\rm Tr}\,\left \{S^i T^a \langle {\bf x}|G|{\bf x}\rangle
S^i T^a \langle {\bf x}|G|{\bf x} \rangle\right \}.
\end{array}
\end{equation}
In this equation ${\bar {\bf B}}$ is the magnetic field associated to the
center ${\bar {\bf A}}$ and $S^i$ is the spin one matrix whose elements
$(j,k)$ are given by $i \varepsilon_{ijk}$.
The notation Tr in equation (\ref{ENERGY}) implies a summation over 
both the color and the spacial indices. For instance,
${\rm Tr}\,\langle {\bf x}|G^{-1}|{\bf x}\rangle 
\,=\sum_{i,a}\,(G^{-1})^{aa}_{ii}({\bf x},{\bf x})$.
Finally, the operator $K$ is the second derivative
of the classical energy with respect to the center ${\bar A}^a_i$. It is
given by
\beq\label{KDEF}
K= (-i {\bf S} \cdot {\bfcalD})^2 - {\bf S \cdot  B},
\eeq
where 
\beq\label{COV} {\cal D}^i \equiv \del^i \,-\,i g\A^i
\eeq
denotes the covariant derivative defined
by the background field $\A^i\equiv \A^i_aT^a$.

In the case of non-Abelian gauge theories, however,
the Gaussian functionals
suffer from a major drawback: they do not satisfy the
requirement of gauge invariance (\ref{GAUSS}). It is thus necessary to
project them onto the subspace of gauge invariant states. This
is achieved by means of the formula
\begin{equation} \label{PROJECTION}
\Psi_0[{\bf A}] = \frac{1}{ {\cal N}} \int {\cal D} [U({\bf x})]\,
\Psi_U[{\bf A}]\,,
\end{equation}
where the functional integration is performed over the unitary 
$N\times N$ matrix field $U({\bf x})$, with the adequate group 
invariant measure, 
and ${\cal N}$ is a normalization factor. The integrand in
eq.~(\ref{PROJECTION}) is the gauge-transform of $\Psi[{\bf A}]$:
\begin{displaymath}
\Psi_U[{\bf A}]= \Psi[U{\bf A}U^+ +
\frac{i}{g} U \mbox{\boldmath $\nabla$ } U^+].
\end{displaymath}
The expectation value $E_P$ of the energy in the projected state 
is given by the following formula
\beq \label{EPROJ}
E_P\,=\,\frac{
\int {\cal D} [U({\bf x})] <\Psi|H|\Psi_U>}{
\int {\cal D} [U({\bf x})] <\Psi|\Psi_U>}\,,
\eeq
which should replace eq.~(\ref{ENERGY}) in practical calculations.
Unfortunately,  eq.~(\ref{EPROJ}) cannot be evaluated in closed form
because the functional integral over the group is not Gaussian \cite{KOVNER}.

In what follows, we shall propose an approximation to eq.~(\ref{EPROJ})
which is inspired from techniques used in nuclear physics to calculate
the zero point rotational energy of deformed nuclei. 
The starting point is the formula
for the collective rotational energy of a classical rotating rigid body    
\begin{displaymath}
E= \frac{1}{2}\, {\cal I}_{ij} \omega_i \omega_j,
\end{displaymath}
where $ {\cal I}_{ij}$ is the inertia tensor and $\omega_i$ the angular
velocity along the i-th axis.
When deformed solutions are obtained in solving 
the nuclear Hartree-Fock equations, the
Thouless-Valatin formalism provides a quantum extension of this formula. It
gives the energy gain when projecting on zero
angular momentum states as \cite{THOULESS,VILLARS}
\begin{equation} \label{DELTA}
\Delta E= \frac{<J_x^2>}{2 {\cal I}_x}+ \frac{<J_y^2>}{2 {\cal I}_y}
+ \frac{<J_z^2>}{2 {\cal I}_z}.
\end{equation}
In this formula $J_x$ is the angular momentum operator along the x axis and
${\cal I}_x$ the corresponding moment of inertia. This quantity is the
polarizability 
\beq
{\cal I}_x = \lim_{\omega \to 0} \frac{<J_x>}{\omega_x}
\eeq
of the system in the presence of an external constraint $\omega_x J_x$.
Equation (\ref{DELTA}) is an approximate expression valid in the limit of a
large fluctuation in the angular momentum $<J^2> \gg 1$ \cite{VILLARS}.

In the case of gauge theories it is straightforward to generalize the above
formula (see \cite{HEINEMAN} for more details).
The moment of inertia is now a matrix in 
color and position in space defined as the polarization tensor
\begin{equation} \label{INERTIA}
{\cal I}^{ab}({\bf x}, {\bf y}) = 
\frac{\delta < {\cal G}^a({\bf x})>}{\delta \omega^b({\bf y})} 
\big\vert_{\omega=0},
\end{equation}
in the presence of an external constraint
\beq
H_{ext}= \int {\rm d}^3 y~\omega^b({\bf y})  {\cal G}^b({\bf y}).
\eeq
In the equations above, ${\cal G}$ is the generator of time -independent
gauge transformations (cf. eq.~(\ref{GL})).
The gain in energy when projecting a wave functional $\Psi[{\bf A}]$ onto
the subspace of gauge invariant states reads, in analogy to
eq.~(\ref{DELTA}),
\beq\label{DETV}
\Delta E= \int {\rm d}^3 x {\rm d}^3 y
\langle \Psi | {\cal G}^a({\bf x}){\cal G}^b({\bf y})| \Psi \rangle
\langle a,\,{\bf x}|\frac{1}{2 {\cal I}}|b,\,{\bf y}\rangle .
\eeq
The expectation values in eqs.~(\ref{INERTIA}) and (\ref{DETV}) are all
Gaussian and can be exactly computed. Nevertheless, the ensuing calculation
of the moment of inertia is extremely tedious, which invites us to look
for a convenient approximation. Namely, we shall assume in what follows 
that the
condensate is large enough so that we can replace the field operator
$A$ by its mean value ${\bar A}$ in the expression (\ref{GL}) for the
gauge generator. We thus define:
\beq\label{GAUSSMF}
{\bar {\cal G}}({\bf x})\,\equiv\,\bfgrad \cdot {\bf E}({\bf x})
\,+\,ig[\A^i,E^i]\,.\eeq
It is then straightforward \cite{HEINEMAN} to obtain the corresponding,
 approximate form of the inertial moment:
\begin{equation} \label{INERTIAMF}
\langle a,\,{\bf x}|{\cal I}|b,\,{\bf y}\rangle \approx\,
\frac{\delta <{\bar {\cal G}}^a({\bf x})>}{\delta \omega^b({\bf y})} 
\vert_{\omega=0}\,=\,
\langle a,\,{\bf x}|\Pi^2|b,\,{\bf y}\rangle\,
\end{equation}
where $\Pi^2$ denotes the square of the kinetic momentum
$\Pi_j\equiv i{\cal D}_j=i\del_j+g{\bar A}_j$.
Furthermore, from eq.~(\ref{GTWF}) and (\ref{GAUSSMF}), one
readily obtains:
\beq
\langle \Psi| {\bar {\cal G}}^a({\bf x})
{\bar {\cal G}}^b({\bf y})| \Psi \rangle\,=\,
\frac{1}{4}\,
\langle a,\,{\bf x}|\Pi_i\, G^{-1}_{ij} \Pi_j|b,\,{\bf y}\rangle\,, \eeq
so that the projection energy  (\ref{DETV}) can be written as:
\beq \Delta E &=&\frac{1}{8}\int {\rm d}^3 x\,
\langle a,\,{\bf x}|\left(\Pi_i\frac{1}{\Pi^2}\Pi_j\right)
  G^{-1}_{ij}|a,\,{\bf x}\rangle\,.\eeq
The Thouless-Valatin correction thus changes the term Tr$(1/8G)$ of
equation (\ref{ENERGY}) into Tr$(Q/8G)$ where 
\beq
Q_{ij}\equiv \delta_{ij}- \Pi_i \frac{1}{\Pi^2} \Pi_j.
\eeq
It is easily checked that $Q^2$=$Q$ and that the action of $Q$ on a
transverse vector field $A^a_i({\bf x})$ (i.e., such that $\Pi_i A_i$=0)
gives back $A_i^a$. We will refer to $Q$ as projector on transverse fields.
We conclude that the Thouless-Valatin correction makes the kinetical part 
of the energy transverse with respect to the mean-field covariant
derivative ${\cal D}_i$.

Further note that the operator $K$ is transverse as well. In fact,
eq.~(\ref{KDEF}) can be rewritten as
\beq
K_{ij}\,=\,\Pi^2\delta_{ij}\,-\,\Pi_i\Pi_j\,+\,2[\Pi_i,\Pi_j]\,, \eeq
which is manifestly transverse: $\Pi_i K_{ij}=0$.
Then, by minimizing
the projected energy $E - \Delta E$ with respect to $G$,
we immediately find that the kernel $G^{-1}$ must be transverse,
\beq\label{TR}
{\cal D}^i\,G^{-1}_{ij}\,=\,0.\eeq
This condition has a simple interpretation: it is equivalent
to ${\bar {\cal G}}({\bf x}) \Psi[{\bf A}]=0$, showing that
the Gauss law (\ref{GAUSS}) is approximately satisfied by the 
Gaussian (\ref{GTWF}).

Eq.~(\ref{TR}) requires the operator $G^{-1}$ to be a functional of the
background field ${\bar A}_i$. A
specific Ansatz for $G^{-1}$ can be obtained from the
variational equation for the projected energy $E - \Delta E$.
To this aim, we write $G_T^{-1}\equiv QG^{-1}Q$ (and similarly
$G_T\equiv QGQ$), and consider the gap equation for $G_T$.
In order to exploit this equation,
we consider a restricted variational space defined by
the following  background field:
\begin{equation} \label{ABAR}
{\bar A}_x=0,~~~~{\bar A}_y=x~T^{3} B,~~~~{\bar A}_z=0. 
\end{equation}
This corresponds to a constant magnetic field in the z-direction and in the
third color with a strength $B$ which is a variational parameter. For
such a background, one can show \cite{HEINEMAN}
that the most general kernel $G_T$ which is consistent with the
gap equation is of the form
\begin{equation} \label{ANSATZ}
\frac{1}{4 G_T^2} \,=\,K\,+\,\beta Q {\bf S \cdot  B} Q =
Q~(\Pi^2 + \alpha  {\bf S \cdot  B}) Q,
\end{equation}
where the dimensionless coefficient $\beta$ = 2+$\alpha$
is a new variational parameter.
The acceptable range of values of $\beta$ is $\beta \ge 1$ \cite{HEINEMAN}.
Indeed, for $\beta < 1$ the previous equation is undefined because
of the occurrence of negative modes in the right hand side
\cite{SAVVIDY,NIELSEN,OLESEN}. 

The expression of the energy density involves quantities such as
$<{\bf x}|G_T|{\bf x}>$ and $<{\bf x}|G_T^{-1}|{\bf x}>$. A proper
definition of these quantities requires a regularization scheme. A
convenient one in the present context is Schwinger's proper time
representation which gives \cite{BREZIN,SCHWINGER}
\beq
\langle {\bf x}|G_T|{\bf x}\rangle=
\frac{1}{\sqrt\pi}
\int_{1/\Lambda^2}^{\infty}
\frac{dt}{\sqrt{t}}
\langle {\bf x}| e^{-t(K+\beta Q  {\bf S \cdot  B} Q)}|{\bf x}\rangle,
\eeq
and
\beq
\langle {\bf x}|G^{-1}_T|{\bf x}\rangle=
\frac{1}{\sqrt\pi}
\int_{1/\Lambda^2}^{\infty}
\frac{dt}{t^{3/2}}\langle {\bf x}|
1-e^{-t(K+\beta Q  {\bf S \cdot  B} Q)}|{\bf x}\rangle.
\eeq
Integrations for small values of the proper time $t$ have been regularized
by the introduction of a cutoff $\Lambda$. In the limit of a large cutoff
one derives
\beq
{\rm Tr} (G^{-1}_T)=\frac{1}{8\pi^2}
\left\{
2 \Lambda^4- 
c_N (\frac{11}{3} -2 \beta + \frac{\beta^2}{3}) 
g^2 B^2 \ln\frac{\Lambda^2}{B}+\ldots 
\right\},
\eeq
and
\beq
\langle {\bf x}|G_T|{\bf x} \rangle=
\frac{1}{16\pi^2}
\left\{ 
\frac{2}{3} \Lambda^2+
(1-\frac{\beta}{3}) {\bf S \cdot  B}
\ln \frac{\Lambda^2}{B}+\langle {\bf x}|G_F|{\bf x} \rangle+ 
{\cal O} (\frac{1}{\Lambda^2})
\right\}.
\eeq
In this last equation $G_F$ is the finite part of $G_T$ in the limit of a
large cutoff. An approximate closed formula for this term can be obtained by
ignoring the projection on transverse states in the Ansatz (\ref{ANSATZ}) 
for the kernel of
the Gaussian. In this case one  finds indeed \cite{HEINEMAN}
\beq \label{GFINI}
\langle {\bf x}|G_F|{\bf x} \rangle = \frac{1}{16 \pi^2} \int_0^\infty 
\frac{ds}{s^2}\left \{ \frac{eBs}{ \sinh eBs}
( e^{es \alpha {\bf S \cdot  B}} - es \alpha  {\bf S \cdot  B}) - 1 
\right\}.\eeq
The resulting energy density reads 
\begin{equation} \label{HBBETA} \begin{array}{lll} 
\langle \Psi|{\cal H}({\bf x})|\Psi\rangle&=&
\frac{1}{32\pi^2} \Lambda^4 + \frac{1}{2}B^2-
\frac{1}{32 \pi^2} c_N
\{ \frac{11}{3} - \frac{\beta^2}{3} \} g^2 B^2 \ln \frac{\Lambda^2}{B} \\
&& \\ 
&+& \frac{6}{(64 \pi^2)^2} c_2^2 c_N^2 B^2 
(1-\frac{\beta}{3})^2 g^4 \ln^2 \frac{\Lambda^2}{B} +
(\frac{g^2}{32 \pi^2})^2 c_2 c_N 
\Lambda^2 {\rm Tr} \langle {\bf x}|G_F|{\bf x} \rangle,
\end{array}
\end{equation}
where $c_N$= Tr($T^{(3)} T^{(3)}$) = N for the group SU(N).
Ignoring the $B$-independent divergent piece $\sim \Lambda^4$,
the remaining expression in eq.~(\ref{HBBETA}) should become finite
after the renormalization of the coupling constant: that is,
$\langle \Psi|{\cal H}({\bf x})|\Psi\rangle$
 must become finite after
sending the bare coupling strength to zero
logarithmically in the limit of a large cutoff. But obviously
this procedure cannot work for the last term
in eq.~(\ref{HBBETA}), which involves a quadratic ultraviolet divergence.
We conclude that the renormalizability of
$\langle \Psi|{\cal H}({\bf x})|\Psi\rangle$ requires the trace of the finite
part of the kernel to vanish identically, which then gives a
condition on the variational parameter $\beta$= 2 + $\alpha$. Using the
approximate expression  (\ref{GFINI}), this condition reads:
\beq 
\int_0^\infty 
\frac{ds}{s^2}\left \{ \frac{eBs}{ \sinh eBs} (1+2 \cosh \alpha eBs) - 3 
\right\}=0.
\eeq
An approximate solution of this equation (with non-vanishing $B$)
can be found by expanding the
hyperbolic functions to second order. One thus obtains:
\beq
\alpha^2 \simeq \frac{1}{2}\,,~~~~{\rm i.e.}~~~~\beta \simeq 
2 \pm \frac{1}{\sqrt{2}}.
\eeq
Remarkably, these values of $\beta$ guarantee that, in the present calculation,
there is no problem with the negative modes \cite{HEINEMAN}
(recall that, for $\beta=0$, one faces
infrared divergences at large proper times  \cite{MAIANI}; see,
e.g., eq.~(\ref{GFINI})).
The minimum in the energy density can be seen to correspond to the value
$\beta=2-1/\sqrt{2}$. For this value of $\beta$,
the energy density is, in the case of the SU(3) gauge group,
\beq 
\langle {\cal H} \rangle=\frac{1}{2}B^2-
\frac{1}{192 \pi^2} (13+4 \sqrt{2}) c_3 g^2 B^2 \ln \frac{\Lambda^2}{B} + 
\frac{g^4}{12 (16 \pi^2)^2} (3+2 \sqrt{2})c_3^2 B^2 \ln^2 \frac{\Lambda^2}{B}, 
\eeq
where only the $B$-dependent terms have been kept.
The minimum in the energy density occurs for $B=B_{min}$ with
\begin{equation} \label{6e14a}
B_{min}=\Lambda^2~~ \exp (-\frac{16 \pi^2}{g^2 c_3} X_{min}), 
\end{equation}
and
\begin{equation}
\label{6e14b}
X_{min}  \simeq 0.363 + \frac{g^2}{32 \pi^2} c_3 + {\cal O}(g^4). 
\end{equation}
The value of the energy at the minimum is
\begin{equation} \label{6e15}
<{\cal H}>_{min} = \frac{B^2_{min}}{2} 
\left\{ (\frac{3}{2} + \sqrt{2}) X_{min} - \frac{13}{4}- \sqrt{2} \right\}
\frac{g^2}{16 \pi^2} \frac{c_3}{3}  \,< 0 . 
\end{equation}
In order to obtain a finite value of the magnetic field at the
minimum when the cutoff $\Lambda$ is sent to infinity, 
namely $B_{min}$= $\mu^2$ / $\sqrt{e}$, 
the coupling constant $g$
must run with the cutoff according to
\beq
\frac{g^2 c_3}{16 \pi^2} \ln \frac{\Lambda^2}{\mu^2}= 0.363.
\eeq
We can now define the effective coupling constant for a given scale $\mu$ by
looking at the coefficient of $B^2$ in the expression of the energy density
and by writing the quantity $\ln (\Lambda^2/B)$ as  
$\ln (\Lambda^2/\mu^2)+\ln (\mu^2/B)$. The result is
\beq 
\frac{1}{g_R^2(\mu^2)}= \frac{1}{g^2}
- \frac{1}{96 \pi^2} (13+4 \sqrt{2}) c_3 \ln \frac{\Lambda^2}{\mu^2} + 
\frac{g^2}{6 (8 \pi^2)^2} (3+2 \sqrt{2})c_3^2 \ln^2 \frac{\Lambda^2}{\mu^2}. 
\eeq

From this equation we obtain the beta function
\beq
\beta(g_R(\mu)) \equiv  \mu \frac{\partial g_R}{\partial \mu}= 
-g_R^3 \frac{1}{48 \pi^2} c_3 \sqrt{52.05},
\eeq
in the limit $g_R^2/g^2 \to$ 0.
It differs from the perturbative one loop result
\beq
\beta_P(g)=-g^3 \frac{1}{16 \pi^2} c_3 \frac{11}{3}.
\eeq
This is not unexpected in
view of the fact that our vacuum state exhibits a non vanishing value of the
condensate and a kernel which largely differs from the perturbative one. In
fact the perturbative result can be recovered in our approach by simply 
ignoring the term $\beta S_z T^{(3)}$ in the trial Gaussian kernel. This
is however not legitimate in the variational context because $\beta$=0
belongs to a region where the kernel exhibits unstable modes.
Our result also differs from that of Kogan and Kovner \cite{KOVNER} 
who use a variational 
Ansatz which does not include the possibility of a condensate.

By looking at the energy expression at the minimum we find that the quantity
$<{\cal H}>_{min}/g^2$ has a limit 
\beq
\frac{<{\cal H}>_{min}}{g^2} \simeq -4.2 \times 10^{-3} \mu^4
\eeq
when the cutoff goes to infinity, i.e. is
a renormalization group invariant quantity. We can compare this result to
the value of the energy density in the bag model $\varepsilon=-B$
where $B$ is now the bag constant \cite{BAG}. With $B$=(240 MeV)$^4$, the
bag model leads to  $\varepsilon$= $-3.3 \times 10^{-3}$ GeV$^4$. This gives
$\mu$=0.941 GeV.

>From our approximate solution for the vacuum functional, it is easy to
calculate the value of the gluon condensate:
\beq
<F_{\mu \nu}^a F_{\mu \nu}^a >=- 2 <E_i^a E_i^a-B_i^a B_i^a>. 
\eeq
At the minimum, we obtain for this quantity a {\it finite} value
\beq
<F_{\mu \nu}^a F_{\mu \nu}^a >= \frac{1}{3}(7 + \sqrt{2}) B_{min}^2 X_{min}, 
\eeq
which gives
\beq
<F_{\mu \nu}^a F_{\mu \nu}^a > \simeq 0.37 \mu^4, 
\eeq
where $\mu \simeq$ 1 GeV. This result is in qualitative agreement with the 
phenomenological value 0.5 GeV$^4$ of the gluon condensate \cite{SHIFMAN}.

{\bf Acknowledgements}. We are most grateful to A. K. Kerman whose remarks
and suggestions have been a continuous source of inspiration during the
course of this work. Stimulating discussions with J. Polonyi are also 
gratefully acknowledged.

\end{document}